\documentclass[aps,floatfix,onecolumn,nofootinbib,12pt]{revtex4}

\usepackage{graphicx}
\usepackage{url}
\usepackage{stfloats}
\usepackage{amsmath}
\usepackage{array}

\begin{document}

	\title{An Update to Isomers of Rydberg Excitations \\in Argon Clusters }
	\author{Mukul Dhiman}
 \affiliation{%
	Université Paris-Saclay, CNRS, Institut des Sciences Moléculaires d’Orsay, \\91405, Orsay,
France
}%
\author{Benoit Gervais}
\affiliation{%
	Normandie Univ, ENSICAEN, UNICAEN, CEA, CNRS, CIMAP, UMR6252, 14000 Caen, France 
}%
\date{\today}

\begin{abstract}
	The effect of Diabatisation is reported in the excited argon isomers using the Diatomic-In-Molecules (DIM) method. In previous work using DIM, the lowest energy isomers of Ar$_N^*$ were shown as Ar$_3^*-$Ar$_{N-3}$, however, using the Hole-Particle-Psedopotential (HPP) method, it was shown that the excitation is localised over dimer not trimer; Ar$_2^*-$Ar$_{N-2}$. In this work we improve the DIM calculations by including previously ignored strongly avoided crossing between 3p4s and 3p4p $^{1,3}\Sigma$ states. 
\end{abstract}
\maketitle
\section{Introduction}
The lowest energy isomers of excited states of excited Argon clusters (Ar$_N^*$) were reported by Naumkin \textit{et al.} \cite{Naumkin1999} almost more than 25 Years ago, where they showed the excitation is localised over trimer. 
Recently, in our previous work \cite{dhiman2022}, using Hole-Particle-Psedopotential (HPP) method\cite{duplaa1996pseudopotential-1,durand1997pseudopotential,Duplaa1996-2}, we showed that the excitation is localised on the dimer rather than the trimer, and the isomers are very different from ionic systems as previously noted. HPP gave us insights into the lowest energy isomers and spectroscopic properties but being computationally expensive, it is not suitable for \textit{on-the-fly} calculations of Potential Energy Curves. 
However, the Diatomic-In-Molecules (DIM) method, which was used to study excited argon clusters previously, is a powerful \textit{on-the-fly} method to study quantum dynamics. Therefore, in the following article, we present an update to the DIM method such that it can be used to study the dynamics of the excited state with a better description of the system. 

In the following work, we first present the DIM set-up to study excited states. Following that, we introduce the diabatisation technique to resolve the strong avoided crossing. And finally, we present our results and compare them to HPP and previous DIM calculations. 

\section{Theory and Method}
The inter-atomic interactions are treated using the Diabatised-Diatomic-In-Molecule (Di-DIM) method with spin-orbit couplings (SOC). The basic method was developed by Ellison \cite{Ellison1963} and others \cite{Tully1973,Tully1976,Tully1977,kuntz1979interaction,Kuntz1979}. The DIM Hamiltonian is given by; 
\begin{equation}
    \hat{H}_{\text{DIM}} = \sum_{A}^M\sum_{B>A}^M \hat{H}_{AB} - (M-2) \sum_A^M \hat{H}_A
\end{equation}
where $\hat{H}_A$ are the atomic fragments, $\hat{H}_{AB}$ are the diatomic fragments and $M$ is the total number of atoms in the polyatomic molecule. 
The wave function for the $l^{\text{th}}$ electronic state solving the electronic Hamiltonian is given as a linear combination of \textit{n}-electron \textit{polyatomic basis functions}(pbf) given as, 
\begin{equation}
\label{eq:wavefxn_DIM}
\Psi^{l}(1,2,\dots,n) = \sum_a c_{a}^{l} \Phi_a^{\text{pbf}}(1,2,\dots,n)
\end{equation}
where $c_{m}^{l}$ are the expansion coefficients for state $l$ associated to pbf $m$.
To define the basis configurations, for the current system which is an excited Argon cluster, all the species are the same with only one of them in an excited state. 
Hence, $\alpha=\beta=\gamma=\dots=$ Ar and we can write the anti-symmetrized pbf for our system as:
    \begin{equation}
        \label{Ar_basis_def}
        \Phi_{a} = A_{i} B_{j} C_{k} \dots 
    \end{equation}
where Ar at position $A$ is in state $i$, Ar at position $B$ in state $j$ and so on. Only configurations like $A_iB_0C_0\dots$, $A_0B_jC_0\dots$, etc are considered. 

The electron of the excited atom is not ionized and is considered to be localised on the same atomic centre as the hole generated to promote the electron from the ground state. For our DIM model, the excited state for an atom is obtained from singly excited configuration $3p^54s$, $3p^54p$, $3p^53d, \dots$. However, in the current study we will consider only the first excited configurations associated to $3p^54s$.
To get the excited state, we start with the following Slater determinant, which is a fair approximation of the $^1S_{1/2}$ ground state of Ar atom, when correlation effects are neglected: 
    \begin{equation*}
        | A_0 \rangle = |3\overline{p}_{-1} 3p_{-1} 3\overline{p}_{0} 3p_{0}3\overline{p}_{1} 3p_{1} \rangle
    \end{equation*}
where $\overline{p}$ is for $\beta$ spin or spin down state. From this reference atomic ground state, an excited state can be written as;
    \begin{equation*}
        \widehat{a}^{\dagger}_x \widehat{a}_a|A_0\rangle \equiv x^{\dagger}a|A_0\rangle\equiv |x^{\dagger}a\rangle
    \end{equation*}
where, $\widehat{a}^{\dagger}_x$ is a creation operator of an electron in spin orbital $x$ and $\widehat{a}_a$ is an annihilation operator of electron in spin orbital $a$ corresponding to a hole creation in orbital $a$. 
For a given atom and given spatial orbital $x$, we have 6-hole possibilities. While taking into account the two spin orientations of the excited orbital $x$ for each configuration we can generate 12 determinants associated with the $3p^54s$ configurations.
Following the steps defined for DIM by Kuntz \textit{et al}\cite{Kuntz1979}, it is essential at this point to adapt our configuration basis on which we will perform rotations. 
They are not ideally suited at this moment as they are not eigenfunctions of $L^2$. A basis set made of such eigenfunctions can however easily be constructed from the above determinant for each shell associated with an orbital $x$. 
Using such an L-adapted basis set will make the rotation algorithm simpler because we can achieve in one single matrix multiplication what would be otherwise done by rotating each orbital, i.e. 6 successive rotations. In the particular case of the determinants $|x^{\dagger}a\rangle$ are proper eigenfunctions of $L^2$. This is generally not the case and $3p^54p$ or $3p^53d$ configurations are combinations of such elementary Slater determinants. We shall thus note the atomic functions accordingly as $A_m = | ^{2S+1}P;M_L;M_S \rangle$, which is sufficient for the $3p^54s$ configurations. The associated configurations are given in supplementary material. \\
The diatomic fragments matrix when represented in a localised pbf basis of atomic functions is given as: \\
\begin{equation}
        \label{eq:deloc}
        \hat{H}_{AB} = \begin{pmatrix}
        h_{A}  &   h_{AB}\\
        h_{BA}  &   h_{B} \\
        \end{pmatrix}
\end{equation}

\begin{align}
    & h_{A} = h_{B} = \frac{1}{2} 
        \begin{pmatrix}
        E_{\Pi_{g+u}^+} & & & \\
        & E_{\Sigma_{u+g}^+} & &\\ 
        & &  E_{\Pi_{g+u}^-} & 
    \end{pmatrix} \\
    & h_{AB} = h_{BA} = \frac{1}{2} 
        \begin{pmatrix}
        E_{\Pi_{g-u}^+} & & & \\
        & E_{\Sigma_{u-g}^+} & &\\ 
        & &  E_{\Pi_{g-u}^-} &
    \end{pmatrix}
\end{align}
where $\Lambda_{u+g} = \frac{1}{2} (\Lambda_u+\Lambda_g)$ and $\Lambda_{g-u} = \frac{1}{2} (\Lambda_g-\Lambda_u)$ for $\Lambda=\Sigma,\Pi$. Therefore, as an example, $E_{\Sigma_{u+g}^+} =  \frac{1}{2} (\Sigma_{u}^++\Sigma_{g}^+) $ and so on. 
We can parametrise the diatomic potential using any \textit{ab initial} method. In previous work \cite{Naumkin1999}, they used PECs obtained using CI \cite{spiegelmann1984calcul}. A detailed comparison of Argon excimer potential can be found here\cite{dhiman2022,dhiman2022modelling}.
In figure \ref{fig:diab_PEC} black solid lines, we show the lowest excited PECs and dotted black lines show the higher excited states. We see the presence of strong coupling ( leading to the hump) associated with $\Sigma_g^+$ 3p4s state. 
Hence, we can not consider this state to be a pure diabatic state, which was previously considered. There is an extensive avoided crossing, therefore an adiabatic to diabatic transformation is essential. 

\subsection{Diabatisation Scheme} 
\begin{figure}[h]
    \includegraphics[scale=0.18]{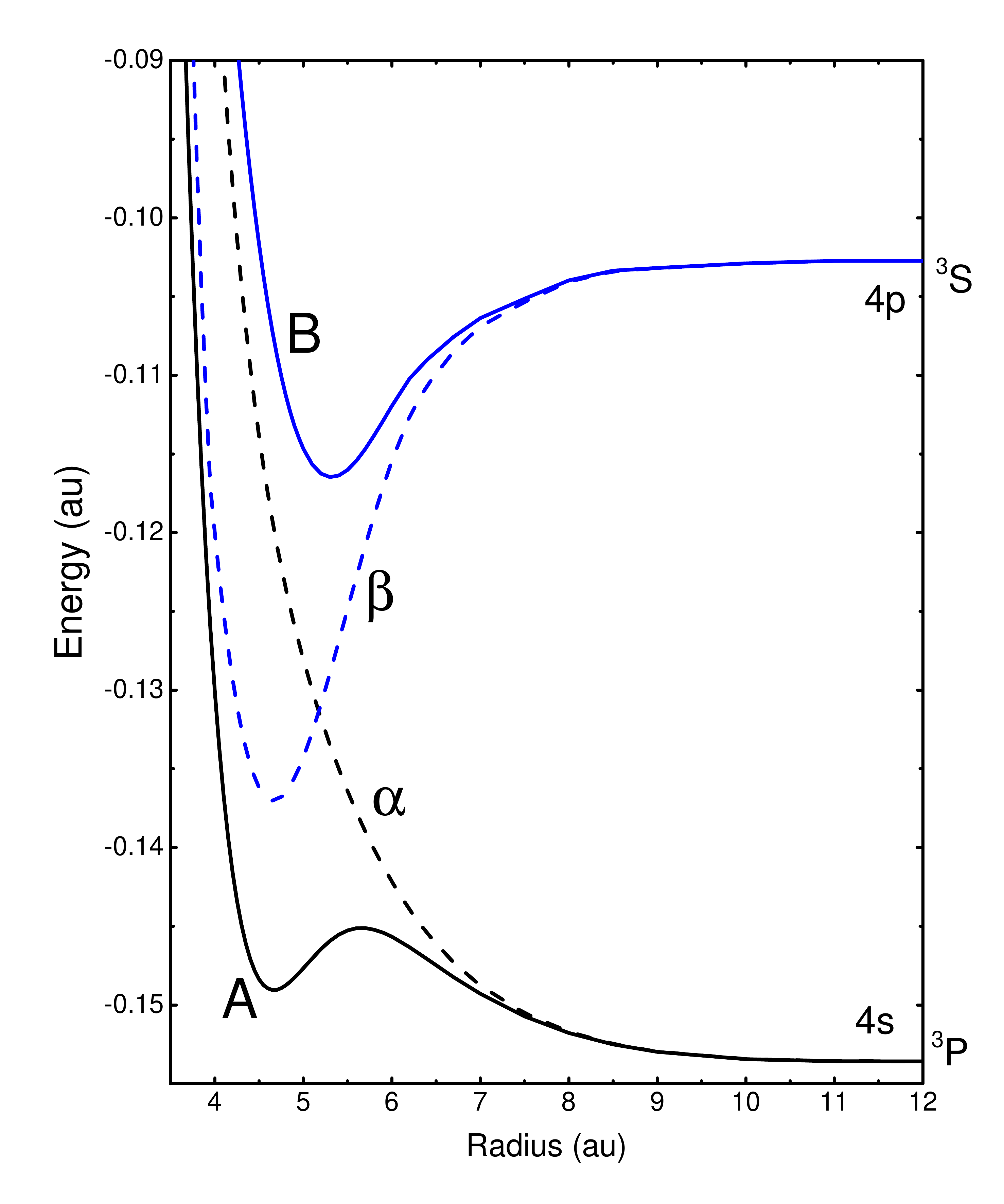}
    \caption{\label{fig:diab_trplet}Diabatised potential-energy curves A ($^{ad}\Sigma_{g}^+$) and B (\textit{ad hoc} state) are the adiabatic states, while $\alpha$ ($^d\Sigma_{g}^+$) and $\beta$ (dummy state) are the diabatic potential curves for the triplet state.\\}
\end{figure}

The uncoupling between two adiabatic states is achieved by unitary transformation \cite{alfalah2010non} which is defined as: 
\begin{equation}
\label{eq:diab_by_rot}
	\begin{pmatrix}
		A & \\
		  & B 
	\end{pmatrix} = U^{\dagger} 
	\begin{pmatrix}
		\alpha & \delta \\
		\delta & \beta 
	\end{pmatrix} U
\end{equation}
where A and B are adiabatic state energy, $\alpha$ and $\beta$ are the diabatic state energy with $\delta$ being the coupling between them.
For the excited Argon system, A state is the adiabatic $^{1,3}\Sigma_{g}^{\text{a}}$, B is an estimated \textit{ad hoc} adiabatic state, using which, we intend to restore the coupling of 3p4s $\Sigma_{g}$ state with the higher states as shown in figure \ref{fig:diab_trplet}. 
$\alpha$ is diabatic $^{1,3}\Sigma_{g}^{\text{d}}$ (obtained using HPP, shown later) and $\beta$ is a dummy diabatic state.
Assuming A, B and $\alpha$ are know, we solve \ref{eq:diab_by_rot} to get the couplings as:
\begin{equation}
\label{eq:diab_couplings}
\delta^2 = (A-\alpha) (\alpha-B)
\end{equation}
and finally, we get the diabatic PEC $\beta$ that is to be used to parametrise DIM Hamiltonian as:
\begin{equation}
    \label{eq:get_diab_beta}
    \beta = \alpha + \delta \left( 1-\frac{1}{\text{tan}\theta} \right)
\end{equation}
where $\theta$ is the angle of rotation. See supplementary material for derivation. \\
Therefore, the Di-DIM Hamiltonian can be given as: 
\begin{equation}
\hat{H}_{\text{Di-DIM}} = 
    \begin{pmatrix}
    h^{D}_A & h^{D}_{AB} \\
    h^{D}_{BA} & h^{D}_{B}
    \end{pmatrix}
\end{equation}  
where $h^{D}_A=h^{D}_B$ and $h^{D}_{AB}$ are arranged as ($A_xA_yA^{\alpha}_zA^{\beta}_z$) for each spin multiplet,
\begin{equation}
    h^{D}_A = \frac{1}{2} \begin{pmatrix}
    E_{\Pi_{g+u}} & & & \\
    & E_{\Pi_{g+u}} & &\\
    & & (\alpha) E_{\Sigma_{u+g}^+} & \delta_{\Sigma_{u+g}}\\
    & & \delta_{\Sigma_{u+g}} & (\beta) E_{\Sigma_{u+g}^+}
    \end{pmatrix}
\end{equation}
and
\begin{equation}
    h^{D}_{AB} =  \frac{1}{2} \begin{pmatrix}
    E_{\Pi_{g-u}} & & & \\
    & E_{\Pi_{g-u}} & &\\ 
    & & (\alpha)E_{\Sigma_{u-g}^+} & \delta_{\Sigma_{u-g}}\\
    & &\delta_{\Sigma_{u-g}} & (\beta) E_{\Sigma_{u-g}^+}
    \end{pmatrix}
\end{equation}
where $\delta_{\Sigma_{u\pm g}}$ are the non-adiabatic coupling terms obtained from diabatization. $E_{\Lambda_{g\pm u}}$ are the energies obtained from the PECs as defined in equation \ref{eq:deloc}.

It is possible to use any method to get the PECs like any DIM model. In this case, we are going to use Ar$_2^*$ PECs obtained using the Hole Particle Pseudopotential (HPP) method discussed in Dhiman \textit{et al} \cite{dhiman2022} and the PhD thesis of Dhiman \cite{dhiman2022modelling}. 
This provides us state $A$ in equation \ref{eq:diab_by_rot} is $^{\text{ad}}\Sigma_{g}^+$, where `ad' is for the adiabatic state as shown in figure \ref{fig:diab_trplet}.
Using the relation between $u$ and $g$ symmetries; 
\begin{equation}
    \Sigma_{u+g}^+ = \frac{1}{2} \left(  ^{\text{d}}\Sigma_{u}^+ +  ^{\text{d}}\Sigma_{g}^+ \right) \rightarrow ^{\text{d}}\Sigma_{g}^+ = 2\times\Sigma_{u+g}^+ - ^{\text{d}}\Sigma_{u}^+
\end{equation}
Using HPP, it is possible to obtain diabatic (eg: $\Sigma_{u\pm g}^+$) PECs. This is done by freezing one of the atoms in the ground state and exploring all the possible states of the excited atom.
This was proposed by Cohen and Schneider to obtain diabatic states of Ne$_2^*$ \cite{cohen1974ground}. These potential curves are obtained in $\Sigma$ symmetry. 
The selection of $B$-state is not trivial. This is an \textit{ad hoc} state, i.e., it is not really present in the Ar$_2^*$ PECs obtained using the HPP method or any other method. Since there is no way of knowing the true nature of coupling or the \textit{ah hoc} state, it is approximated by comparing the lowest energy isomer obtained for trimer while varying the depth of \textit{ah hoc} state. The B-state is selected to dissociate at $4p$ $^3S$ limit as shown. Figure \ref{fig:diab_trplet} shows the diabatized potential curves for triplet. The change in the dissociation limit does not alter the results, provided it is high enough.

Note: Initial work by Cohen and Schneider on Neon \cite{cohen1974ground} assumed the \textit{ah hoc} state to be $np$ excited orbital only and they observed a barrier in the diabatic curve whereas we do not observe any barrier after the minima as seen in $\beta$ curve of figure \ref{fig:diab_trplet}. 
However, the reason for considering an \textit{ah hoc} state is, just by considering $4p$ state (i.e., by just plugging in the 4p states), we were not able to reproduce the lowest isomer of the trimer. 
We also note that with the introduction of 3p4p $\Sigma$ states, our configuration space is increased from 12 to 16 and we neglect the SOCs between the \textit{ad hoc} state in this work for simplicity. 

\subsection{Energy Minimisation and Properties}
Similar to previous work, we limit our search to the lowest excited state energy equilibrium geometries which are associated with the triplet states present under the singlet states. 
In DIM method, it is not difficult to obtain the energy derivatives analytically, and we used them to perform damped molecular dynamics. 

The starting geometries to determine the isomers are selected to be either ground states geometries, or the HPP lowest energy isomer \cite{dhiman2022} or previous isomers reported by Naumkin \textit{at al} \cite{Naumkin1999}. 
With the information of the minimum geometries, we characterise the associated electronic structure by looking at the hole distribution. 
As DIM is based on atomic configuration, using the information of the expansion coefficients, $c_l^m$, we can determine the amount of hole on each atom. 
The electronic distribution is more difficult and we do not expect to obtain faithfully the electronic densities using DIM. 

\section{Results and Discussion}

\begin{figure}[h]
	\begin{center}
		\includegraphics[scale=0.15]{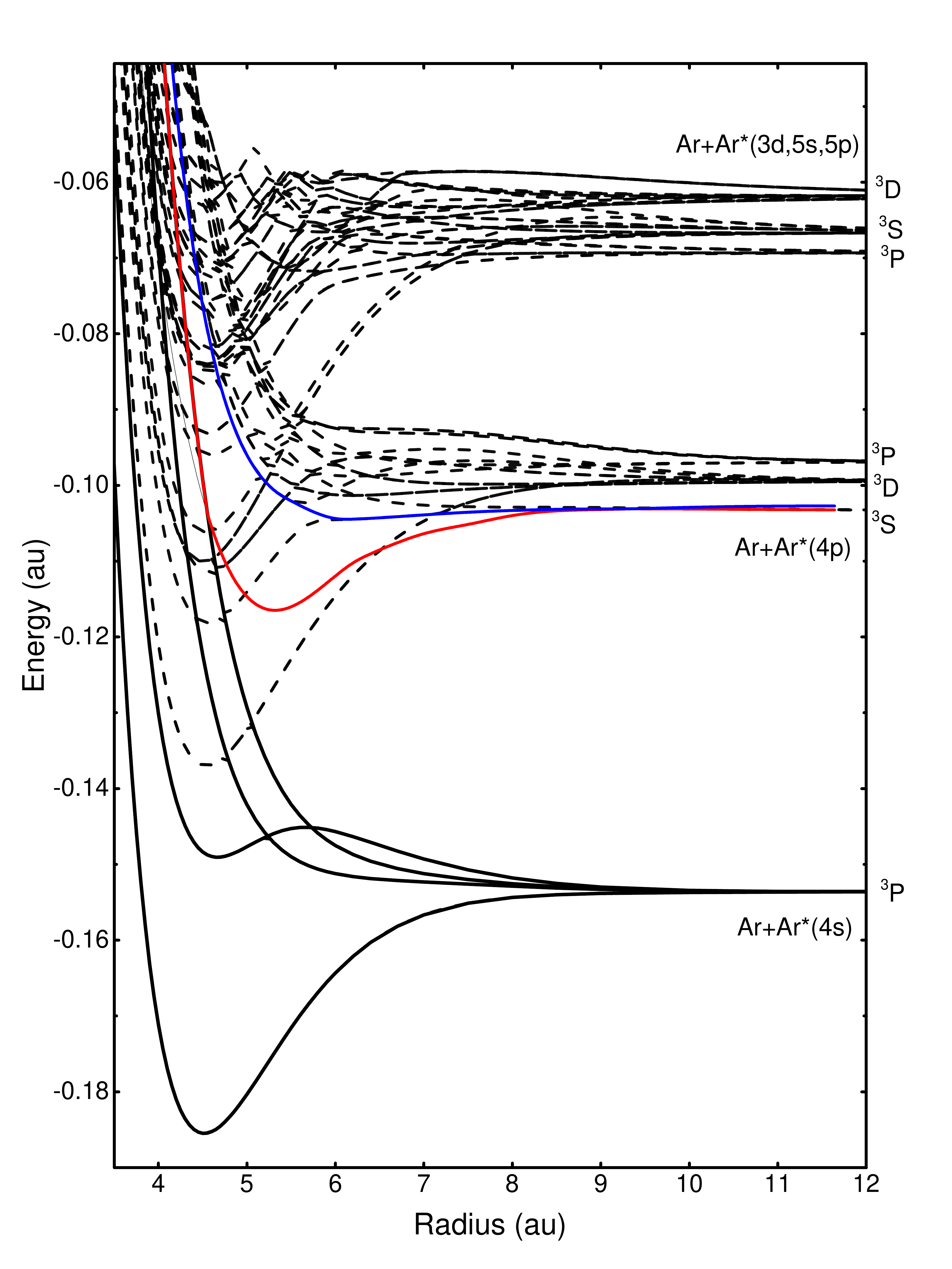}
		\caption{\label{fig:diab_PEC} The excimer PECs calculated using HPP with blue $(2)^3\Sigma_u^+$ and red curves $(2)^3\Sigma_g^+$ showing the \textit{ad hoc} states used to restore the diabatic characteristic of $\Sigma^+$ state.} 
	\end{center}
\end{figure}

Building on the foundational work of Naumkin and Wales \cite{Naumkin1999}, we delve into the four lowest potential energy curves for singlets and triplets, namely $^{3,1}\Sigma_{u,g}^+$ and $^{3,1}\Pi_{u,g}$, as depicted in Figure \ref{fig:diab_PEC}. 
These curves are derived from HPP excimer results. However, as highlighted earlier, these PECs lack true diabatic characteristics. 
To address this, we introduced an additional higher state that emulates the 4s-4p $^{3,1}\Sigma_g$ couplings, effectively resolving the avoided crossing. 
This higher-energy state in the 4p limit is illustrated by the red (g) and blue (u) curves in Figure \ref{fig:diab_PEC} for triplets. 
Notably, the lowest-energy PECs bear a strong resemblance to those obtained through the HPP approach.

The work on the excited argon trimer to get the lowest energy isomer and possible PES was previously done using the DIM method and was limited to 4s configurations \cite{Naumkin1999,goubert1995semiclassical,naumkin2002diatomics}. 
In their approximation, they considered all the lowest states to be purely diabatic.
The lowest energy isomer for Ar$_3^*$ using the DIM method was reported to be symmetric, of D$_{\infty \text{h}}$ symmetry, with the excitation delocalized over all three argon atoms, with higher concentration on the central atom. However, in our previous work using HPP method we have shown that the symmetry of the trimer is lowered to C$_{\infty\text{h}}$ \cite{dhiman2022}. We explore this discrepancy further using DIM with diabatisation. 

\begin{figure}[h]
		\includegraphics[scale=0.80]{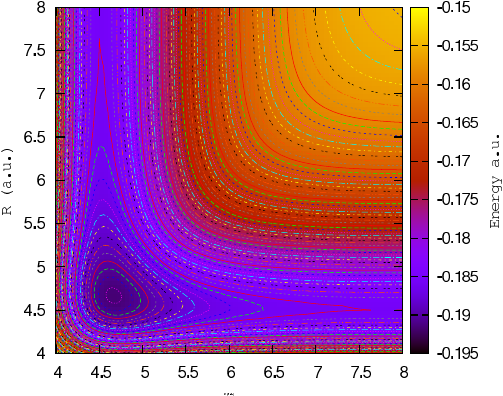} 
        \includegraphics[scale=0.80]{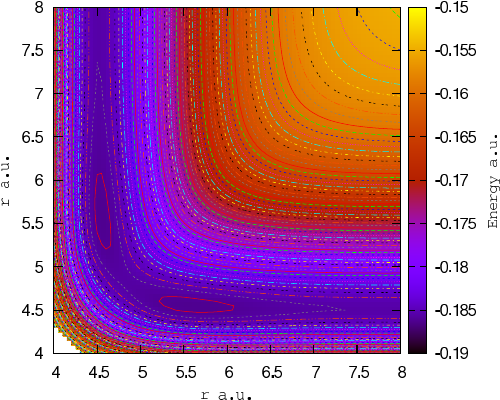}
		\caption{\label{fig:trimer_DIM_colin} Top: Ar$_3^*$ surface plot without diabatization forming the minima for linear-symmetric in C$_{2 \text{v}}$ symmetry. Bottom: Ar$_3^*$ surface plot with diabatization forming the minima for linear asymmetric in D$_{\infty h}$ symmetry.}
\end{figure}

The trimer potential energy surface (PES), with (left) and without (right) diabatization is shown in figure \ref{fig:trimer_DIM_colin} respectively for linear geometry. 
For consistency, the same dimer PECs are used in both methods to parametrize the DIM. 
It is clear that without diabatization we obtain the minimum in D$_{\infty h}$ symmetry, as reported by Naumkin \cite{Naumkin1999} and Goubert \cite{goubert1995semiclassical}. 
With diabatization, we obtain minimum with C$_{\infty \text{h}}$ symmetry similar to HPP. Hence, we show that the DIM with diabatisation (Di-DIM) proves better transferability.

The dissociation energy of the lowest trimer energy observed using Di-DIM is different from the one observed in HPP in spite of the linear anti-symmetric geometry observed through the two methods. 
For the Ar$_2^*-Ar$ trimer, the dissociation energy is 160 cm$^{-1}$ with Di-DIM and 290cm$^{-1}$ with HPP, both lower than the CASPT2 value of 340cm$^{-1}$. 
However, these values are not as high as the 1300cm$^{-1}$ obtained using DIM without diabatization. The Di-DIM trimer results align better with the HPP model isomers, but the excimer-ground state atom distance is shorter in Di-DIM (5.60 au) compared to HPP (6.10 au) and CASPT2 (5.90 au). 

\begin{figure}[h]
		\includegraphics[scale=0.72]{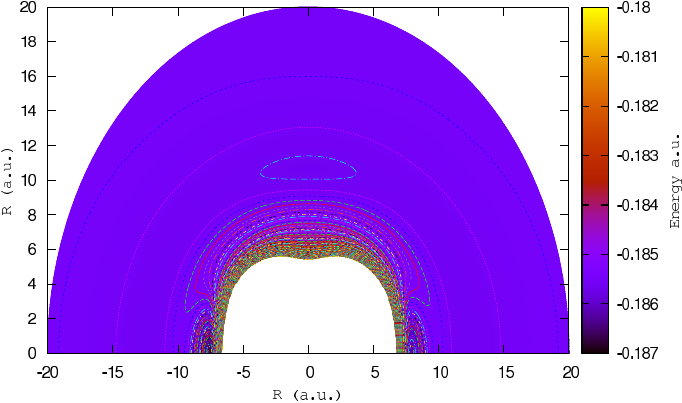}
        \caption{\label{fig:trimer_DIDIM_Rot} Ar$_3^*$ surface plot in C$_{2 \text{v}}$ symmetry while rotating the ground state atom around excimer using Di-DIM.}
\end{figure}

We demonstrate that employing an \textit{ad hoc} potential energy curve (PEC) enables us to replicate the influence of higher states within the Di-DIM method. However, defining the nature of this PEC is challenging because there is no predefined function available to generate such a curve. Consequently, the \textit{ad hoc} state presented here represents the best approximation achievable within the current scope of our work.
It is crucial to highlight that the choice of the \textit{ad hoc} state does not significantly impact the inter-atomic distance or the dissociation energy. Variations in the dissociation limits of different \textit{ad hoc} states result in changes to the trimer dissociation energy by approximately 10-20 cm$^{-1}$, which is relatively insignificant. Similarly, the corresponding inter-atomic distance changes by about 0.002 atomic units (au). 
When examining the C$_{2 \text{v}}$ symmetry, we observe a saddle point formation in the HPP model that is absent in the Di-DIM method. Specifically, the T-shaped geometry in C$_{2\text{v}}$ symmetry, which appears as a saddle point in the HPP model, is identified as a local minimum in the Di-DIM method. This observation aligns with previous findings using the DIM approach, as illustrated in Figure \ref{fig:trimer_DIDIM_Rot}. 
This discrepancy is likely attributable to the rotational effects of the $P$ and $D$ states, which differ from those of the $S$ state and are not accounted for in our current Di-DIM model. These findings suggest that incorporating additional higher states, along with their respective couplings, is necessary to achieve results that are more consistent with HPP calculations.

Although the trimer results are not sufficient, they are consistent with the HPP and CASPT2 observations. We now explore the bigger clusters and discuss the effect of diabatization and its importance in the lowest energy isomers.

\begin{figure} [h]
		\includegraphics[scale=0.7]{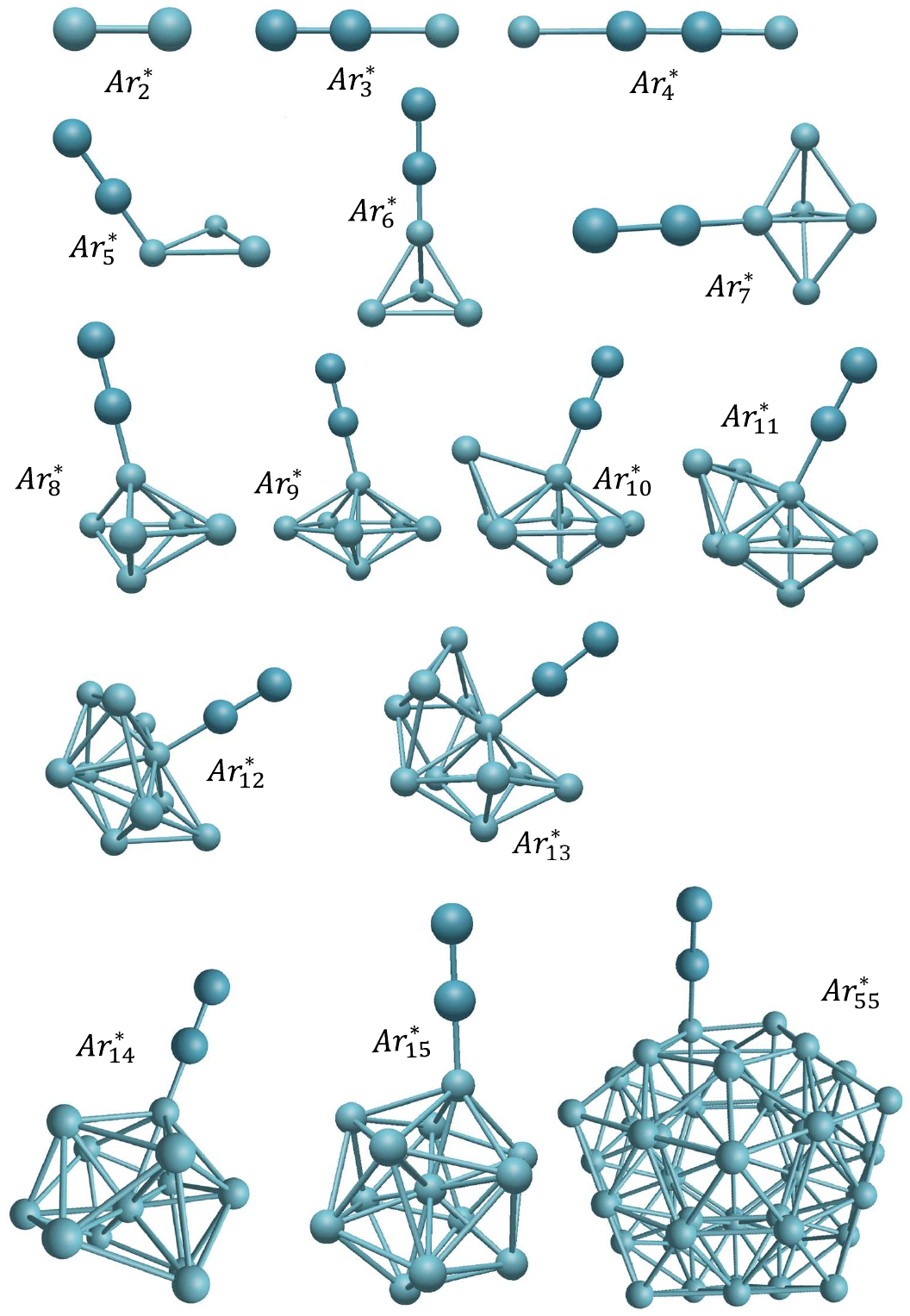}
		\caption{\label{fig:isomers_MD_dim}   All isomers obtained using DIM with diabatization for Ar$_N^*$ for $2\leq N \leq 15$ and $N = 55$. The size of the excimer is represented by enlarging the Argon excimer. }
\end{figure}

Damped molecular dynamics is used to obtain the lowest energy structures in the Di-DIM method, which are summarised in Figure \ref{fig:isomers_MD_dim}. 
Generally, we observe an excimer attached to the (N-2) ground state cluster, similar to the trend seen in HPP. 
This contrasts with the previous DIM method, where an excited trimer is attached to the ground state cluster. In previous DIM calculations, 80\% of the excitation (hole) was reported to be located on the central atom of the excited trimer, while the remaining 20\% was shared by the other two atoms. In our Di-DIM method, the total excitation (hole) is equally distributed between the two outermost atoms, consistent with HPP calculations.

The geometry of the lowest energy trimer isomer serves as the foundation for most larger isomers, closely aligning with results from both HPP and DIM. Although the Di-DIM model cannot replicate the dipole observed in HPP, the use of diabatization in conjuncture with the approximate \textit{ad hoc} PECs allows us to obtain geometries of the lowest energy isomers similar to those from HPP. Notably, the distance of the excimer in most geometries observed using Di-DIM is shorter compared to those derived from HPP calculations.

The Di-DIM dimer characteristics are identical to the one obtained using HPP method, as \textit{a priori}, we parameterise the Di-DIM dimer potential using HPP with the addition of \textit{ad hoc} PEC. Therefore, the excimer inter-atomic distance $R_e=4.52$ au. 

As reported in lowest energy isomers obtained using HPP method, for $N=4$ the lowest energy isomer is observed to be the one where the excimer is located between two ground state atoms.  
However, unlike in the HPP model, the two neutral atoms are not equidistant from the excimer. This discrepancy is likely due to the absence of polarization effects in the DIM method and the specific potential energy curves (PECs) chosen.
Interestingly, this isomer is not observed without diabatization in DIM. 

\begin{figure}[h]
	\begin{center}
		\includegraphics[scale=0.30]{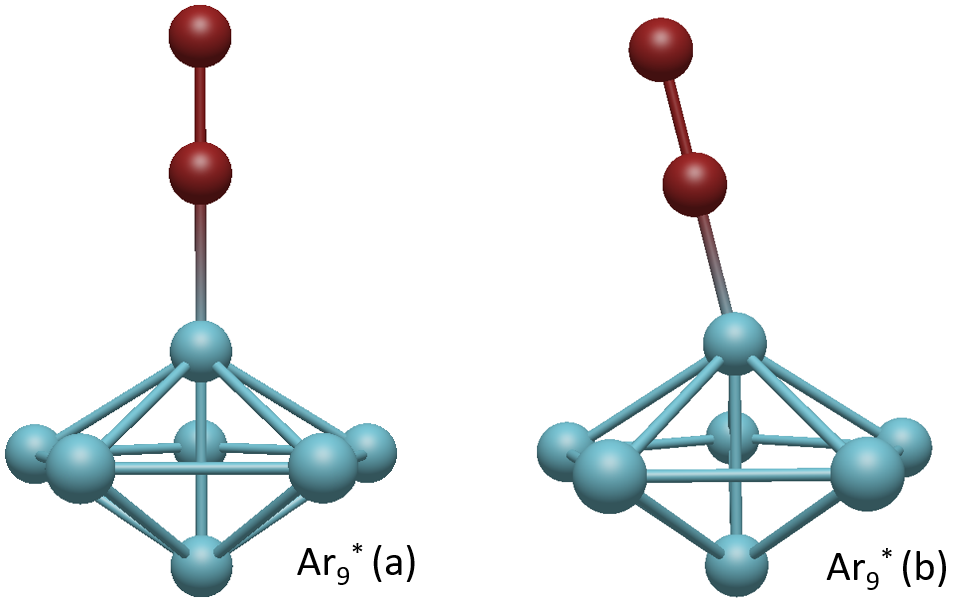}
		\caption{\label{fig:isomer_Ar9_di-dim}   This figure shows the difference between the geometry for the excited argon cluster of N=9 for HPP and Di-DIM. Figure (a) shows the isomer obtained using HPP and (b) shows the isomer obtained using Di-DIM which is tilted showing symmetry breaking.}
	\end{center}
\end{figure}

We consider now an example of $N=9$ whose lowest-energy isomer geometry is very similar using HPP and DIM methods. In Di-DIM, the lowest energy isomer has an excimer attached at a longer distance than the excimer distance found in HPP calculations systematically concerning Ar$_3^*$. 
This isomer also breaks the rotational symmetry along the excimer: the excimer is tilted in one direction as shown in Figure \ref{fig:isomer_Ar9_di-dim}. The reasoning is similar to symmetry breaking observed in $N=4$, i.e. lack of polarization in the DIM method. 

The Di-DIM results are summarised in table \ref{TABISOMER-DiDIM} for cluster sizes up to N=15, supplemented by the cluster size N=55 corresponding to a rare-gas magic number. 
The dissociation energies obtained using Di-DIM ($^AD_e^*$) are compared to HPP dissociation energies ($^BD_e^*$), where D$_e^*=\frac{(E_N^*-E_2^*)}{(N-1)}$. With the ground state dissociation energy $D_e$. 

As expected, the dissociation energy increases with the increasing cluster size as it reaches the continuum limit. We see that the HPP dissociation energies are much higher than the ground-state dissociation energy, and contrarily the Di-DIM dissociation energies are lower than the ground-state dissociation limit(with the exception to $N=4$, which is closer to HPP). 
However, they are not lower than the single ground state atom dissociation.

It is difficult to reach the bulk dissociation limit with a cluster size of $N=15$ as the maximum number of bonds per atom (i.e. 6) which corresponds to half of the cohesion energy ($\approx300$ cm$^{-1}$) of an argon cluster with 12 bonds per atom ($\approx 600$ cm$^{-1}$). Similarly, for $N=55$, there are more surface states than the bound states. A cluster with more than $N\approx10^{4}$ atoms would reach a limit where the bound states are more than the surface states. 

The Di-DIM geometries of the lowest energy isomers are shown in figure \ref{fig:isomers_MD_dim} for $2\leq N \leq 15$ and $N=55$. The gerometry of the lowest energy isomers is a mixture of isomers derived from HPP, and DIM by Naumkin while others are somewhere in the middle of the two. Similar to HPP, all the excitation is localised on the dimer resulting in the geometry where an excimer is attached to $N-2$ ground state cluster. The only exception is for $N=4$, where the excimer is positioned between the two ground state atoms, akin to the HPP model.

\section{Conclusion}
    In the following work we studied lowest energy isomers of Ar$_N^*$ using Diabatised-DIM for $N=2-15$ and $55$. We do a direct comparison with previous work by Naumkin \textit{et al} \cite{Naumkin1999} using DIM. In that work they neglected the strong avoided crossing which was thought to not play significant role in lowest energy isomers where they reported the excitation to be shared by a timer on the surface. 
    In our previous work using HPP method, we showed that in the lowest energy isomers, the excitation is shared by an excimer. To determine the discrepancy between the two results, in the work, we focused on the strong avoided crossing in $\Sigma$ state. 
    Here, using diabatisation, we uncouple the strongly coupled lowest 3p4s sigma states from 3p4p using an \textit{ad hoc} approximation of 3p4p $^{1,3}\Sigma$ state. We show, even by using a crude approximate state, the primary character of the lowest excited isomers are in better agreement with the HPP results. 

    While exploring the lowest energy isomers with Di-DIM we observe that the excitation is localised on the protruding dimer, as observed in the lowest-energy isomers obtained using the HPP model; Ar$_2^*-$Ar$_{N-2}$. 
    However, the geometry associated with the lowest energy isomers is not the same as obtained using the HPP model, but they are similar. 
    We further observed symmetry breaking of several HPP lowest energy isomers, where the excimer attached to the ground state cluster is observed to be tilted. 
    The isomers from Di-DIM are observed to be mixtures of both previous DIM and HPP calculations. We also observed the exception for $N=4$ in Di-DIM, like for HPP isomer. 
    However, the Di-DIM isomer for Ar$_4^*$ is linear asymmetric rather than D$_{\infty h}$ symmetric like HPP. This isomer does not exist in previous DIM calculation. 
    
    The Di-DIM results are limited due to the lack of possibility to include dipole-moment and higher excited states. However, these are sufficient to study excited state dynamics where we primarily focus of localisation of excitation. 
    In our next work, we will show how dynamics changes if these couplings are ignored. 

\section*{Disclosure statement}
No potential conflict of interest was reported by the author(s).

\bibliographystyle{unsrt}
\bibliography{paper}

\newpage
\section{Appendix: Supplementary Material}
\subsection{Application to $3p^54s$ excited configuration}

To get $3p^54s$ excited state, we put $x^{\dagger} = 4s$, so we can write this triplet $^3P$ excited state as:
\begin{equation}
    |C,{4s^{\dagger}},{}^3P;1,1\rangle
\end{equation}
where $C=3p^54s$, $M_L=1$ and $M_S=1$.

\begin{equation*}
|C,{4s^{\dagger}},{}^3P;1,1\rangle = |\overline{3p}_{-1}4s^{\dagger}\rangle = |4s 3p_{-1} \overline{3p}_{0} 3p_{0}\overline{3p}_{1} 3p_{1} \rangle
\end{equation*}
$\overline{3p}_{-1}$ is spin down. Now to reduce the angular momentum, we apply $L^-$ as follows, 
\begin{equation}
\label{L-}
L^-|C,{}^3P;1,1\rangle = \sqrt{L(L+1)-M_L(M_L-1)}|C,{}^3P;0,1\rangle
\end{equation}
it can be shown that the application of $L_-=\sum_i l_i^-$ as a determinant is equivalent to the application of $\sum_k l_k^-$, where $k$ runs over the orbitals in the determinant.
\begin{align*}
& \sum_{k}l_k |4s 3p_{-1} \overline{3p}_{0} 3p_{0}\overline{3p}_{1} 3p_{1} \rangle \\ & = \sqrt{1(1+1)-0(0-1)}|4s 3p_{-1} \overline{3p}_{-1} 3p_{0}\overline{3p}_{1} 3p_{1} \rangle
\end{align*}
on rearranging the orbital ordering we get, 
\begin{equation*}
|C,{}^3P;0,1\rangle=-|\overline{3p}_{-1}3p_{-1} 4s 3p_{0}\overline{3p}_{1} 3p_{1} \rangle = -\mid\overline{3p}_{0}4s^{\dagger}\rangle 
\end{equation*}
apply $L_-|C,{}^3P;0,1\rangle$ we get, 
\begin{equation*}
|C,{}^3P;-1,1\rangle=|\overline{3p}_{-1}3p_{-1} \overline{3p}_{0}3p_{0}4s3p_{1} \rangle = \mid\overline{3p}_{-1}4s^{\dagger}\rangle 
\end{equation*}
Hence we get the transformation matrix elements as, 
\begin{equation}
T_{4s}= \frac{1}{\sqrt{2}}
\begin{pmatrix} 0 & 0 & 1 \\ 0 & -1 & 0\\1 & 0 & 0 \end{pmatrix}
\begin{pmatrix} & \overline{3p}_{1}4s^{\dagger} \\ &\overline{3p}_{0}4s^{\dagger} \\ &\overline{3p}_{-1}4s^{\dagger}  \end{pmatrix}
\end{equation}

\subsection{Application to $3p^54p$ excited configuration}

To get $3p^54s$ excited states, we put $x^{\dagger} = 4p_{1}$, so we can write this $^3D$ excited state as $ |C,{4p_{1}^{\dagger}},{}^3D;2,1\rangle $ where $C=3p^54p$, $M_L=2$ and $M_S=1$.
\begin{equation*}
\mid C,{4p_{1}^{\dagger}},{}^3D;2,1\rangle = |\overline{3p}_{-1}4p_{1}^{\dagger}\rangle = |4p_{1} 3p_{-1} \overline{3p}_{0} 3p_{0}\overline{3p}_{1} 3p_{1} \rangle
\end{equation*}
Now to reduce the angular momentum, we apply $L^-$ as in equation \ref{L-}, 
\begin{align*}
L^-|C,{}^3D;2,1\rangle &= \sqrt{2(2+1)-2(2-1)}|C,{}^3D;1,1\rangle \\ 
&= \sqrt{4}|C,{}^3D;1,1\rangle 
\end{align*}
\begin{multline*}
\sum_{k}l_k |4p_{1} 3p_{-1} \overline{3p}_{0} 3p_{0}\overline{3p}_{1} 3p_{1} \rangle = \sqrt{1(1+1)-1(1-1)} \\
\lbrace |4p_{0} 3p_{-1} \overline{3p}_{-1} 3p_{0}\overline{3p}_{1} 3p_{1} \rangle +\mid4p_{1} 3p_{-1} \overline{3p}_{-1} 3p_{0}\overline{3p}_{1} 3p_{1} \rangle\rbrace
\end{multline*}
after ordering the orbitals and using simplified notations we get, 
\begin{equation*}
|C,{}^3D;1,1\rangle = \dfrac{1}{\sqrt{2}} \lbrace\vert\overline{3p}_{-1}4p_0^{\dagger}\rangle-\vert\overline{3p}_{0}4p_1^{\dagger}\rangle\rbrace
\end{equation*}
In the same spirit, on using $L_-$ operator we can get all the following configuration, 
\begin{equation*}
|C,{}^3D;0,1\rangle = \dfrac{1}{\sqrt{6}} \lbrace\vert\overline{3p}_{-1}4p_{-1}^{\dagger}\rangle-2\vert\overline{3p}_{0}4p_0^{\dagger}\rangle+\vert\overline{3p}_{1}4p_1^{\dagger}\rbrace
\end{equation*}
\begin{equation*}
|C,{}^3D;-1,1\rangle = \dfrac{1}{\sqrt{2}} \lbrace\vert\overline{3p}_{1}4p_0^{\dagger}\rangle-\vert\overline{3p}_{0}4p_{-1}^{\dagger}\rangle\rbrace
\end{equation*}
and  
\begin{equation*}
|C,{}^3D;-2,1\rangle = \vert\overline{3p}_{1}4p_{-1}^{\dagger}\rangle
\end{equation*}
To write the $^3P$, we orthogonalize to $^3D$ states of same $M_L$ and $M_S$, for the uppermost $M_L$ in the P-shell, i.e. $M_L=1$ and the lowest $M_L=-1$,
\begin{equation*}
|C,{}^3P;1,1\rangle = \dfrac{1}{\sqrt{2}} \lbrace\vert\overline{3p}_{-1}4p_0^{\dagger}\rangle+\vert\overline{3p}_{0}4p_1^{\dagger}\rangle\rbrace
\end{equation*}
\begin{equation*}
|C,{}^3P;-1,1\rangle = \dfrac{1}{\sqrt{2}} \lbrace\vert\overline{3p}_{1}4p_0^{\dagger}\rangle+\vert\overline{3p}_{0}4p_{-1}^{\dagger}\rangle\rbrace
\end{equation*}
By using $L_-$ on $|C,{}^3P;1,1\rangle$ we get, 
\begin{equation*}
|C,{}^3P;0,1\rangle = \dfrac{1}{\sqrt{2}} \lbrace\vert\overline{3p}_{-1}4p_{-1}^{\dagger}\rangle-\vert\overline{3p}_{1}4p_1^{\dagger}\rangle\rbrace
\end{equation*}
and the $^3S$ state by orthogonalization to both $|C,{}^3D;0,1\rangle$ and $|C,{}^3P;0,1\rangle$, 
\begin{equation*}
|C,{}^3S;0,1\rangle = \dfrac{1}{\sqrt{3}} \lbrace\vert\overline{3p}_{1}4p_{1}^{\dagger}\rangle+\vert\overline{3p}_{0}4p_0^{\dagger}\rangle+\vert\overline{3p}_{-1}4p_{-1}^{\dagger3}\rangle\rbrace
\end{equation*}

\subsection{Diabatization}
\label{appen:sec_diabat}
\begin{equation}
    \begin{pmatrix}
    A & \\
      & B
    \end{pmatrix} 
    = U^{\dagger} 
    \begin{pmatrix} \alpha & \delta \\ \delta & \beta \end{pmatrix} 
    U 
\end{equation}
where $U$ is the rotation matrix defined as: 
\begin{equation}
    U = \begin{pmatrix} \cos{\theta}  & - \sin{\theta}  \\  \sin{\theta}  & \cos{\theta} \end{pmatrix}
\end{equation}
Therefore we have: 
\begin{align}
       \begin{pmatrix} A & \\ & B \end{pmatrix} &= 
       \begin{pmatrix} \cos{\theta}  &  \sin{\theta}  \\ - \sin{\theta}  & \cos{\theta} \end{pmatrix} 
       \begin{pmatrix} \alpha & \delta \\ \delta & \beta \end{pmatrix} 
       \begin{pmatrix} \cos{\theta}  & - \sin{\theta}  \\  \sin{\theta}  & \cos{\theta} \end{pmatrix} \\
       &= \begin{pmatrix} \cos{\theta}  &  \sin{\theta}  \\ - \sin{\theta}  & \cos{\theta} \end{pmatrix} 
       \begin{pmatrix} \alpha \cos{\theta} +\delta \sin{\theta}  & -\alpha \sin{\theta} +\delta \cos{\theta}  \\ 
       \delta \cos{\theta} +\beta \sin{\theta}  & -\delta \sin{\theta} +\beta \cos{\theta}  \end{pmatrix} \\
       &= \begin{pmatrix} 
       \alpha \cos^2{\theta}+2\delta \cos{\theta}  \sin{\theta} +\beta \sin^2{\theta} &  \delta( \cos^2{\theta}- \sin^2{\theta})+ \sin{\theta}  \cos{\theta} (-\alpha+\beta) \\ 
       \delta( \cos^2{\theta}- \sin^2{\theta})+ \sin{\theta}  \cos{\theta} (-\alpha+\beta) & \alpha \sin^2{\theta}-2\delta \cos{\theta}  \sin{\theta} +\beta \cos^2{\theta} \end{pmatrix}
\end{align}
For $\delta( \cos^2{\theta}- \sin^2{\theta})+ \sin{\theta}  \cos{\theta} (-\alpha+\beta)=0$ we have: 
\begin{align}
    \delta( \sin^2{\theta}- \cos^2{\theta}) &= (\beta-\alpha) \sin{\theta}  \cos{\theta}  \\
    \beta-\alpha &= \frac{\delta( \sin^2{\theta}- \cos^2{\theta} )}{ \sin{\theta}  \cos{\theta} } \\
    \beta = \alpha &+ \delta \left( 1-\frac{1}{\tan{\theta}} \right) 
\end{align}

\begin{table}[h!]
\begin{center}
  \caption{  Characteristics of Ar$_N$ clusters in their lowest excited triplet state. The comment indicates whether the geometry is similar to the geometry obtained by relaxation with the Di-DIM model. $^AD_e^*$ (cm$^{-1}$) and $^BD_e^*$ (cm$^{-1}$) are the dissociation energies of Di-DIM and HPP respectively.}
  \label{TABISOMER-DiDIM}
  \begin{tabular}{ c c c c c c c}
     \hline
    isomer & Symmetry               & Energy (au) & $^{\text{A}}D_e^*$  & $^{\text{B}}D_e^*$ & $D_e$ & comment \\
    \hline
    2    &   D$_{\infty {\rm h}}$   & -0.18549 & 0   & 0 & 50  &   - \\
    3    &   C$_{\infty}$           & -0.18620 & 79  & 147 & 99 &   HPP    \\
    4    &   D$_{\infty {\rm h}}$   & -0.18766 & 159 & 164 & 149 &  HPP    \\
    5    &   C$_{\infty}$           & -0.18826 & 153 & 195 & 180 &  N-DIM  \\
    6    &   C$_{\infty}$           & -0.18997 & 197 & 221 & 202 &  N-DIM  \\
    7    &   C$_{\infty}$           & -0.19120 & 209 & 232 & 231 &  N-DIM  \\
    8    &   C$_{\infty}$           & -0.19278 & 229 & 253 & 243 &         \\
    9    &   C$_{\infty}$           & -0.19430 & 242 & 270 & 262 &  Both    \\
    10    &   C$_{\infty}$          & -0.19539 & 242 & 278 & 277 &  Both    \\
    11    &   C$_{\infty}$          & -0.19743 & 263 & 297 & 290 &  Both    \\
    12    &   C$_{\infty}$          & -0.19946 & 280 & 311 & 307 &         \\
    13    &   C$_{\infty}$          & -0.20514 & 360 & 316 & 330 &  Both    \\
    14    &   C$_{\infty}$          & -0.20428 & & - & 331 &  N-DIM  \\
    15    &   C$_{\infty}$          & -0.20685 & 335 & - & 337 & N-DIM  \\
    55    &   C$_{\infty}$          & -0.30584 & 487 & - & 477 &  -       \\
    \hline
  \end{tabular}
\end{center}
\end{table}

\end{document}